\documentclass[pra,aps,twocolumn,showpacs,superscriptaddress,amsmath,amssymb]{revtex4-1}
\usepackage{color}

\usepackage{amsmath,graphicx}

\begin{document}
\def\tr{\rm{Tr}}
\def\la{{\langle}}
\def\ra{{\rangle}}
\def\a{{\alpha}}
\def\e{\epsilon}
\def\q{\quad}
\def\w{\tilde{W}}
\def\t{\tilde{t}}
\def\a{\hat{A}}
\def\h{\hat{H}}
\def\E{\mathcal{E}}
\def\p{\hat{P}}
\def\u{\hat{U}}
\def\n{\hat{n}}
\def\j{\hat{j}}
\def\r{\hat{\rho}}
\def\s{\hat{S}}
\def\.{\cdot}
\def\lef{\leftarrow}

\title{Path integral approach to space-time probabilities: a theory without pitfalls but with strict rules
}
%
%
\author {D.  Sokolovski}
\affiliation{Departmento de Qu\'imica-F\'isica, Universidad del Pa\' is Vasco, UPV/EHU, Leioa, Spain}
\affiliation{IKERBASQUE, Basque Foundation for Science, E-48011 Bilbao, Spain}

\date{\today}
\begin{abstract}
Following the renewed interest in the topic \cite{HY}, we revisit the problem of assigning probabilities to classes of Feynman paths passing through specified space-time regions.
We show that by assigning of probabilities to interfering alternatives, one already makes an assumption that  the interference has been destroyed through interaction with an environment, or a meter. 
Including the effects of the meter allows to construct a consistent theory, free of logical 'pitfalls' identified in Ref.\cite{HY}. Wherever a meter cannot be constructed, or cannot be set to effect the desired decoherence, formally constructed probabilities have no clear physical meaning, and can 
violate the necessary sum rules. We illustrate the above approach by analysing the three examples considered in \cite{HY}.

\end{abstract}

%
%
\maketitle
%
%
%
%
%
%
%
\section{Introduction}
Recently Halliwell and Yearsley \cite{HY} revisited the problem of assigning probabilities 
to amplitudes obtained by restricting Feynman paths to space-time regions.
They emphasised that the method suffers from serious deficiencies: "seemingly obvious notion of 'restricting paths' leads to the quantum Zeno effect and hence to unphysical results" 
This is the sense in which we would say that path integrals constructions may suffer from pitfalls'\cite{HY}.
Their criticism extends to the path integral analysis of the quantum traversal time \cite{TT1}-\cite{TT7}. In order to mitigate the Zeno effect, the authors of \cite{HY}
follow Alonso {\it et al} \cite{KICK} in suggesting 'temporal coarse graining' whereby the observed system is controlled only at discrete times, between which it  is allowed to evolve freely. The problem is closely related to the more general question concerning the origin of quantum probabilities (see, for example,  \cite{PR1}-\cite{PR6}). 

We disagree with the conclusions of Ref.\cite{HY} with regards to the 'pitfalls' of the path integral approach to quantum probabilities, and would like to restore, in as much as possible,  the good name of quantum traversal time. The purpose of this paper is to show that the approach is indeed a consistent theory, with its own strict rules, yet free from 'unphysical' features.
We base our argument on a simple consideration which, it appears, was not taken into account in Ref.\cite{HY}.
Assigning probability to interfering alternatives, one must assume that the said interference has somehow been destroyed. Destruction of interference is a physical process, which must result from interaction with another physical system, e.g., a meter.  We demonstrate that once the inevitable effects of the meter are properly taken into account,  one has a theory free from logical contradictions.
The rest of the paper is organised as follows. In Section II we briefly describe path decomposition of the propagator. In Section III  we choose a variable, and construct its amplitude distribution by restricting Feynman paths. In Sect. IV we show that the chosen restriction automatically prescribes the type of the meter required to destroy interference. In Sect. V we emphasise the distinction between the finite-time and continuous quantum measurements, and proceed to analyse the former type. In Sect. VI we formulate three questions which define a quantum measurement.
Section VII describes the mixed state of the system after interference between the classes of Feynman paths has been destroyed. In Section VIII we prove a general result concerning the emergence of a kind of the Zeno effect in high-accuracy 'ideal' measurements.
In Sections IX, X, XI we analyse the examples used in Ref.\cite{HY} to illustrate 'unreasonable properties' of path integrals amplitudes. Section XII contains our conclusions.



\section{Path decomposition of a transition amplitude}
For a system governed by a Hamiltonian $\h$, consider the transition amplitude 
\begin{eqnarray}\label{0}
K(F,I,t)=\la\psi_F|\exp(-i\h t)|\psi_I\ra
\end{eqnarray}
 between some initial and final states $|\psi_I\ra$ and $|\psi_F\ra$. 
In its Hilbert space, choose a compete orthonormal  basis $|x\ra$ where $x$ may take discrete or continuos values, so that ($I$ is the unity)
\begin{eqnarray}\label{1}
\sum_x |x\ra\la x| = I.
\end{eqnarray}
Writing $\exp(-i\h t)$ as $\prod_{k=1}^K\exp[-i\h(tk/K)]$, inserting (\ref{1}) between the exponentials and also before and after $|\psi_I\ra$ and $|\psi_F\ra$, and sending $K$ to infinity yields the celebrated Feynman path integral \cite{FEYN} (or, as the case may be, path sum), 
\begin{eqnarray}\label{2}
K(F,I,t)=\sum_{paths} A[path].
\end{eqnarray}
Here a path is defined by the sequence of the $x$'s labelling the states through which the system passes at the intermediate times $t'=0,\e,2\e,...,t$, $\e\equiv t/K$. In the continuous limit $K\to \infty$ we will denote it $x(t')$
The functional $A[path] \equiv lim_{K\to \infty}\la\psi_F|x_K\ra\la x_K|\exp(-i\h\e)|x_{K-1}\ra$
$...\exp(-i\h \e)|x_0\ra\la|x_0|\psi_I\ra$ is the amplitude the path contributes to $K(F,I,t)$. 
With Feynman paths and their contributions $A[path]$ known, the task of evaluation of the transition amplitudes reduces, at least formally, to addition of complex numbers. Conceptual simplicity of the Feynman quantum mechanics is matched by  the difficulty of actually performing the path sum (\ref{2}). It does, however,  make a convenient starting point for a discussion of quantum measurements.

\section{Measurable properties of  a quantum system}

Next we may wish to enquire about the system's behaviour in the time interval between its preparation in the state $|\psi_I\ra$ and its subsequent detection in $|\psi_F\ra$.
We may not be interested in all the details, but just in the values of some quantity $F$. For the answer to exist, the value of $F$ must be defined for each of the possible histories, i.e.,  should be a functional $F[path]$ defined on the Feynman paths. It is reasonable to re-arrange the paths according to the value of $F$, and define the probability amplitude to have this value equal to $f$ as
  ($\delta(z)$ is the Dirac delta)
\begin{eqnarray}\label{a1}
\Phi(F,I,t|f)=\sum_{paths}A[path] \delta(F[path]-f).
\end{eqnarray}
Equation (\ref{a1}) can be also written in an operator form
\begin{eqnarray}\label{a2a}
\Phi(F,I,t|f)=\la\psi_F|\u(t|f)|\psi_I\ra
\end{eqnarray}
where the restricted evolution operator propagates the initial state only along the paths which have the property $F[path]=f$.
If there are several quantities $F_1$, $F_2$, ...$F_N$, the probability amplitude for them to have, jointly, the values $f_1$,$f_2$...$f_N$, $K(F,I,t|f_1,f_2,..,f_N)$, can be constructed in a similar manner.
With the part of the path summation (\ref{2}) already performed, the full propagator is given by an ordinary quadrature,
\begin{eqnarray}\label{a2}
K(F,I,t)=\int df_1\int df_2...\int df_N\Phi(F,I,t|f_1,f_2,..,f_N).\q
\end{eqnarray}

It is tempting to define the probability to have the property $F[path]=f$ simply as 
\begin{eqnarray}\label{a3}
P(F,I,t|f)\to |\Phi(F,I,t|f)|^2.
\end{eqnarray}
Next we will show that this temptation should so far be resisted.

\section{Path restriction vs. dynamical interaction}
The task of converting the amplitudes (\ref{a1}) into probabilities requires some additional care.
Our re-arrangement of the Feynman histories into classes was purely cosmetic. Like the paths themselves, the classes remain interfering alternatives. 
To assign the probabilities one must first destroy the interference. This must be done by bringing the system in contact with another quantum system, or systems, and we must decide what type of an additional system (a meter) is suitable for the task.
\newline 
Conveniently, the answer is already contained in our choice of the functional $F$.
Consider the equation of motion satisfied by the restricted path integral (\ref{a1}). If this equation in maps onto a Schroedinger equation describing the original system {\it plus} another system, we immediately obtain a recipe for constructing the meter with the desired properties. If, on the other hand, the equation of motion does not look like such a Schroedinger equation, we must stop and admit that the interference cannot be destroyed by any means available to us. 
The question whether further progress can be made in this latter case is beyond the 
scope of this paper.
\newline
The type of the equation satisfied  by the restricted propagator depends on the choice of the functional $F$, and must be established in each individual case. One general result was proven in Refs.\cite{DS1}: let the functional be of the form
\begin{eqnarray}\label{b1}
F[path]=\int_0^t\beta(t') a(x(t')) dt'
\end{eqnarray}
where $\beta(t)$ is a known function of time and $a(x)$ is some function of $x$. 
Then the probability amplitude for the system to arrive at a 'location' $|x\ra$ by taking only the paths
with the properties $F[path]=f$, $\Phi(x,t|f)$ (we drop the implied dependence in the initial state $|\psi_I\ra$ to simplify notations), satisfies the Schroedinger equation (SE)
\begin{eqnarray}\label{b2}
\partial_t\Phi=[\h-i\partial_f\beta(t) \a]\Phi
\end{eqnarray}
with the initial condition
\begin{eqnarray}\label{b3}
\Phi(x,0|f)=\la x|\psi_I\ra \delta(f)
\end{eqnarray}
Here the operator $\a$, diagonal in the chosen representation, is defined by the choice of $a(x)$ in Eq.(\ref{b1}),
\begin{eqnarray}\label{b4}
\a\equiv \int dx |x\ra a(x) \la x |.
\end{eqnarray}
Equation (\ref{b2}) is, of course, a SE describing the system interacting with a von Neumann meter \cite{vN}, 
with pointer position $f$, designed to measure the operator $\a$.
Unlike in the original von Neumann's approach, the pointer remains coupled to the system for a finite time, since the quantity $F$ in Eq.(\ref{b1}) refers to the whole of the time interval $[0,t]$.
\section{Finite-time measurements vs. their continuous counterparts}
Before proceeding we must emphasise an important distinction between two kinds of measurements.  Suppose we have only one meter, and in the spirit of \cite{HY}, \cite{KICK}, activate it 
through a sequence of sharp strong pulses, that is we choose
 \begin{eqnarray}\label{cb1}
\beta(t')=\sum_{n=1}^N\delta(t-t_n), \q t_n=nt/(N+1)
\end{eqnarray}
Assume, for simplicity, that the operator $\a$ in Eq. (\ref{b4}) is a projector onto a part of the Hilbert space $\Omega$. Then the pointer would move one notch to the right each time the system is in $\Omega$ at $t=t_n$, or otherwise  remains where it is. At the time $t$ we look at the pointer once, and, having found it shifted by $m$ notches, conclude that the system was in $\Omega$ $m$ times out of all $N$ trials. We cannot, however, say exactly when, as the outcome of the measurement is a single number, yielding the value of a functional on otherwise unspecified virtual trajectories. Since the observation took $t$ seconds to complete, we call it a {\it finite-time} measurement.

Suppose that instead we have $N$ identical meters, only one of which is briefly activated at each $t=t_n$, that is, we choose
 \begin{eqnarray}\label{cb2}
\beta_n(t')=\delta(t-t_n),
\end{eqnarray}
As  a result, at $t$ we have $N$ readings, all either $0$ or $1$, and the exact knowledge at which of the $t_n$ 
the system was found in $\Omega$.
Accordingly, there is  a set of probabilities 
$P(x_1,x_2,..x_N)$, $x_n=\pm1$. This is a prototype of a {\it continuous measurement} \cite{MENS}, whose outcome is a real trajectory followed by the observed system. 
Like the authors of \cite{HY} we are interested in finite-time measurements, which we will consider through the rest of the paper. 
\section{Three questions to a quantum measurement}
We see now that to describe a quantum measurement, one must provide clear answers to at least three following questions:

a) What is being measured?

b) How it is being measured?

c) To what accuracy is it being measured?
\newline
To answer the first question, we need to specify a variable which characterises the system in the absence of a meter. In our case it is the functional in (\ref{b1}), whose physical meaning depends on the choice of the switching function $\beta(t')$. Thus, for $\beta(t') =1/t = const$ it represents the time average of the of the quantity $\a$, and for $\beta(t')=\delta(t'-t_0)$, the instantaneous value of $\a$.
Choosing $\beta(t')=\delta(t'-t_1)\pm \delta(t'-t_2)$ allows to measure the sum or the difference of $A$'s values at $t_1$ and $t_2$, and so forth.
\newline
To answer the second question one must ask first if there is a suitable meter can be found.
For a functional of the type (\ref{b1}) the answer is 'yes'.
 One should then prepare the system in the desired initial state, set the pointer to zero, turn on the interaction, and accurately measure the pointer's position at the time $t$.
\newline
The third question follows from the second. The amplitude $\Phi$ in Eq.(\ref{b2}) is not normalisable due to the presence of the $\delta(f)$ in (\ref{b3}), and cannot be used to construct physical probabilities. To obtain a physical amplitude $\Psi(x,0|f)$, one should prepare the pointer in a physical state $G(f)$, 
$\int df |G(f)|^2 =1$, and replace the initial condition (\ref{b3}) with
\begin{eqnarray}\label{bb1}
\Psi(x,0|f)=\la x|\psi_I\ra G(f).
\end{eqnarray}
The result can be written in an equivalent form \cite{DS1}
\begin{eqnarray}\label{bb2}
\Psi(x,t|f)=\int df G(f-f')\Phi(x,t|f').
\end{eqnarray}
which has a simple interpretation. 
 The function $G(f-f')$ plays the role of a filter, selecting a limited range of the values of $F[x(t)]$'s which contribute to the pointer's advancement to position $f$. For an accurate measurement one should choose $G(f)$ sharply peaked around zero, with a small yet finite width $\Delta f$. Now 
 \begin{eqnarray}\label{bb3}
P(x,t|f)=|\Psi(x,t|f)|^2
\end{eqnarray}
yields the probability the find the system in $|x\ra$, and the pointer at a location $f$.
It is also the probability that the {\it observed} system arrives at $x$ and $F$ has the value in the interval  $[f-\Delta f, f+f-\Delta f]$. Accordingly, 
 \begin{eqnarray}\label{bb4}
P(t|f)=\int dx P(x,t|f)
\end{eqnarray}
yields the probability to find the pointer at $x$ without asking about the state of the system.
It is also the probability that, for the observed system, $F$ has the value within $[f-\Delta f, f+f-\Delta f]$ regardless of where it ends up once the measurement is finished.
The correct normalisation of $P(x,t|f)$,
 \begin{eqnarray}\label{bb5}
\int dx df  P(x,t|f) =1
\end{eqnarray}
 is guaranteed, since the evolution according to the SE (\ref{b2}) is unitary.
We note that the limitation on the accuracy of a measurement is of purely quantum nature:
 the values inside the interval $[f-\Delta f, f+f-\Delta f]$ cannot be distinguished since interference between them has not been destroyed.
\section{State of the system after a measurement}
We note that restricting its evolution to paths with the property $F[path]=f$, we effectively 'chop' the state of the system into sub-states
 \begin{eqnarray}\label{c1}
|\Phi(t|f)\ra = \int dx |x\ra \Phi(x,t|f). 
\end{eqnarray}
which add up to the Schroedinger state of a freely evolving system,
 \begin{eqnarray}\label{c2}
 \int df| \Phi(x,t|f)\ra =\exp(-i\h t)|\psi_I\ra. 
\end{eqnarray}
 This is the particular property of the interaction in Eq.(\ref{b2}). It can be seen as the quantum analog of the condition that a classical meter should monitor the measured system without affecting its evolution (for details see Refs.\cite{DS2}). A meter of a finite accuracy $\Delta f$, 
uses linear combinations of  the fine-grained sub-states (\ref{c1})
 \begin{eqnarray}\label{c3}
|\Psi(t|f)\ra = \int df' G(f-f')|\Phi(t|f)\ra,
\end{eqnarray}
and then destroys the coherence between these 'coarse-grained' states. The final mixed state of the system $\r$ after measurement, therefore, is 
 \begin{eqnarray}\label{c4}
\r(t) =\int df |\Psi(t|f)\ra\la\Psi(t|f)|,
\end{eqnarray}
with $tr \r(t)=1$, as follows from Eq.(\ref{bb5}).
\section{Possibility of a 'Zeno effect'}
The convolution formula (\ref{b3}) has the advantage that if the main properties of the fine-grained amplitude $\Phi(x,t|f)$ are known, we can qualitatively estimate the result of smearing it with 
the function $G(f)$. For example, suppose that a transition is classically allowed. Than, in the semiclassical limit, rapidly oscillating $\Phi(x,t|f)$ has a narrow stationary region around the classical value of $F[x(t)]$, $f_{cl}$. Then a not-too-accurate meter will always return the value
$f_{cl}$, since if $G(f-f')$ is centred at  any $f\ne f_{cl}$ the integral (\ref{b3}) will vanish, destroyed by the oscillations \cite{DS3}. 

It is possible to make another general statement about the properties of $\Phi(x,t|f)$.
{\it For a system confined to a finite interval (volume) $a\le x \le b$, the amplitude distribution $\Phi(x,t|f)$ cannot be a smooth function for all $x$. 
Rather, to ensure conservation of probability in the high accuracy limit $\Delta f\to 0$,
it must have singularities, typically, of the Dirac delta type.}

The proof is based on conservation of probability. Suppose we have some reference function $G(f)$ and want to improve the accuracy by making it narrower. This can be achieved by simple scaling, 
 \begin{eqnarray}\label{dd1}
G(f)\to G_{\alpha}(f) = \alpha^{1/2}G(\alpha f),
\end{eqnarray}
where we recalled that $G(f)$ is also the initial state of the pointer, and as such must be normalised to unity, $\int df|G_{\alpha}(f)|^2 =1$. Suppose now that $\Phi(x,t|f)$ is smooth.
Then, as $\alpha \to \infty$, the width of $G_{\alpha}\sim \Delta f/\alpha \to 0$, an we should be able 
to take $\Phi$ outside the integral (\ref{b3}), 
 \begin{eqnarray}\label{dd2}
\Psi(x,t|f)=\int df G_{\alpha}(f-f')\Phi(x,t|f') \approx
\\ \nonumber
 \Phi(x,t|f)\int G_{\alpha}(f')df'
=\alpha^{-1/2}C \Phi(x,t|f),
\end{eqnarray}
where $C=\int G(f')df'$.
In the high-accuracy (ideal measurement)  limit $\alpha \to 0$, the r.h.s. of Eq.(\ref{dd2}) vanishes, and with it vanishes the probability to find the pointer at any location $f$,  $P(t|f)=\int_a^b|\Psi(x,t|f)|^2 dx$, which is of course wrong. The way around this difficulty is to assume that in addition to its smooth part $\tilde{\Phi}(x,t|f)$, 
$\Phi(x,t|f)$ has a number of $\delta$-singularities, 
\begin{eqnarray}\label{dd3}
\Phi(x,t|f)=\tilde{\Phi}(x,t|f) +\sum c_k(x)\delta(f-f_k).
\end{eqnarray}
Now as the accuracy improves, the probability $P(x,t|f)$ becomes
\begin{eqnarray}\label{dd4}
lim_{\alpha \to \infty }P(x,t|f)=\q\q\q\q\q\\ \nonumber
 \sum_k |c_k(x)|^2 \delta (f-f_k) + \alpha^{-1}|C|^2 |\tilde{\Phi}(x,t|f)|^2.
\end{eqnarray}
Suppose now that one is conducting an experiment on a large number of identical systems, all post-selected in the state $|x\ra$, and receives a signal proportional to the number of cases in which the pointer is found in $f$. As the accuracy improves, the signal is dominated by strong peaks at $f=f_k$. In addition, there is a smaller signal revealing more and more details of 
the structure of $|\tilde{\Phi}(x,t|f)|^2$,  and eventually fading altogether.
The positions of the peaks $f_k$, and the coefficients $c_k$ must be determined for each particular case.

For example, earlier we have shown \cite{DS4} that for a system in a finite-dimensional Hilbert space at attempt to determine precisely the value of the time average of an operator $\a$ inevitably leads to a Zeno effect trapping the system in the eigenstates (eigen sub-spaces of $\a$). Another example of such behaviour was given in Ref.\cite{DS5} which analysed a measurement of qubit's residence time with a slightly more sophisticated variant of the von Neumann meter.
\newline
For a system in an infinite volume, e.g., for $-\infty < x < \infty$, there exists another possibility. While $|\Psi(x,t|f)|^2$ must vanish as $\a \to \infty$, the range of $x$' s involved in the integral 
(\ref{bb4}) may increase proportionally, so that the probability is conserved without $\Phi(x,t|f)$ acquiring singular terms. Physically this would mean that the meter scatters the observed system into a wide range of its final positions. 

For the following, it suffices to note that, as it follows from Eq.(\ref{dd4}), an accurate determination of the value of any quantity of the type (\ref{b1}) may  suppress some of the transitions and lead to sharply defined values $F[path]=f_k$ in those transitions which survive. With some hesitation we follow the authors of Ref.\cite{HY} in calling it a 'Zeno effect' also in the case the measured system has a continuous spectrum. In a conventional Zeno effect \cite{Z0}-\cite{Z1}, frequent observations
trap the system in one  of the eigenstates of the measured quantity. Here, the action of he meter restricts the system to a particular type of evolution, without freezing it altogether.  With this we are ready to analyse the cases discussed by the authors of \cite{HY}.

\section{The probability not to enter the right half-space}
We start with the task of defining the probability that, in one dimension, a free particle of a mass $M$ does not enter the region $\Omega
\equiv 0\le x < \infty$ \cite{HY}. Equivalently, one can ask: 'what is the probability that the particle spends in $\Omega$ precisely a zero duration, $\tau=0$?' The functional yielding the duration a Feynman paths spends in $\Omega$ is well known \cite{TT1},
 \begin{eqnarray}\label{d1}
t_{\Omega}(t,[x(\.)]) = \int_0^t \theta_{\Omega}(x(t')) dt',
\end{eqnarray}
where $\theta_{\Omega}(z)= 1$ for $z$ inside $\Omega$, and zero otherwise. The operator in Eq.(\ref{b3}) is just the projector onto the right half-space, $\a=\p_{\Omega}=\int_0^{\infty}dx|x\ra\la x|$. 
The coupling $-i\partial_{\tau}\theta_{\Omega} (x)$ allows to identify the meter as a continuous version of the Larmor clock \cite{L1}-\cite{DS6}, a large magnetic moment which precesses in a magnetic field for as long as particle remains inside $\Omega$.
The solution to Eq.(\ref{b2}) is given by the Fourier integral \cite{TT5}
 \begin{eqnarray}\label{d2}
\Phi(x,t|\tau)=(2\pi)^{-1}\int_{-\infty}^{\infty} dV\exp(iV\tau)\psi_V(x,t)\\ \nonumber
\psi_V(x,t) \equiv \la x| \exp[-i\h_Vt)|\psi_I\ra,\q\q
\end{eqnarray}
where [we use $ \theta(x)$ for $\theta_{[0,\infty)}$]
 \begin{eqnarray}\label{d3}
\h_V=p^2/2m+V\theta(x).
\end{eqnarray}
In other words, to find the traversal time amplitude distribution one needs to know the results of evolving the initial state for all potential steps added in the right-half space - even though we are discussing the properties of a free particle. The transmission ($T$) and reflection ($R$) amplitudes for such a step at an energy $E$ are easily found to be ($k=\sqrt{2M E}$)
 \begin{eqnarray}\label{d4}
T(k,V)=2/[1+(1-V/E)^{1/2}], \q R=T-1,\q\q
\end{eqnarray}
Now if the initial state is a wave packet, initially in the left half space, and moving from left to right,
 \begin{eqnarray}\label{d4a}
\la x|\psi_I\ra=\int dk A(k)\exp[ikx-iE(k)t],  \q\\ \nonumber
 \la x|\psi_I\ra\equiv 0 \q \mbox{for} \q x\ge 0\q\q
\end{eqnarray}  
it evolves into 
 \begin{eqnarray}\label{d5}
\psi_V(x,t) = \int dk T(k,V) A(k)\exp[ikx-iE(k)t] 
\\ \nonumber
  -\int dk A(k)\exp[-ikx-iE(k)t]\\ \nonumber
+\int dk T(k,V) A(k)\exp[-ikx-iE(k)t].
\end{eqnarray}     
Here the first term is the transmitted part, the second term describes the reflection from an infinite step, $V=\infty$, and the third reflected term accounts for the fact that the step isn't, after all, infinite.
We will consider $t$ so large,  that the transmitted and reflected wave packets are well separated and do not overlap.
With the help of Eq.(\ref{d2}) the amplitude distribution for the duration $\tau$ spent by a free particle in the right half-space  becomes [cf. Eq.(\ref{d3})]
 \begin{eqnarray}\label{d6}
 \\ \nonumber
 \Phi(x,t|\tau) = \left\{ \begin{array}{ll}
         \tilde{\Phi} (x,t|\tau) & \mbox{for \q $x>0$};\\
        -\delta(\tau) \psi_{0}(-x,t) +\tilde{\Phi}(-x,t|\tau) & \mbox{for \q $x<0$}.\end{array} \right. 
  \end{eqnarray}    
Here $ \psi_{0}(x,t)\equiv \la x| \exp[-i\h t)|\psi_I\ra$ is just the freely propagating wave packet,
and $\tilde{\Phi} (x,t|\tau)$ is a smooth function, involving the Fourier transform of $T(k,V)$ at all relevant energies. The last two terms clearly correspond to the particle being reflected, for example,
we have
 \begin{eqnarray}\label{d7}
 -\psi_{0}(-x,t)\equiv \psi_{\infty}(x,t),
\end{eqnarray}
where $ \psi_{\infty}(x,t)$ is the wave packet reflected by an infinite potential wall, $V=\infty$.
This may again seem strange, since everything said so far referred to a free particle.
\newline
The following point is  central to our analysis: 
{\it  assignment of probabilities to interfering classes of quantum histories cannot be considered 
outside the context of measurements of the property of interest.} 
\newline
So far we have only contemplated a classification of Feynman paths according to the duration spent in $\Omega$, yet the result already contains a reference to the reflection which would be caused by a Larmor clock, should we decide to employ one. Moreover, the fine-grained amplitude $\Phi(x,t|\tau)$ contains information about all measurement scenarios, which are, in turn, 
specified by the choice of the filter $G$ in Eq.(\ref{c3}).  
Suppose, for example, we decide to make no measurement at all by making $G(f-f')$ in Eq.(\ref{b3}) so broad that it can be replaced by a constant. It is useful to note that (since $(2\pi)^{-1}\int d\tau 
\int dV exp(iV\tau) T(k,V)=T(k,0)=1$) $\tilde{\Phi} (x,t|\tau)$ add up to the freely propagating pulse, 
 \begin{eqnarray}\label{d7a}
\int_0^t \tilde{\Phi} (x,t|\tau) d\tau= \psi_{0}(x,t).
\end{eqnarray}
Thus with no measurement make the last two terms in (\ref{d6}) cancel, leaving us, as it should, with only the free wave packet travelling to the right of the origin $x=0$.

Or we may want to know what is the probability for spending {\it precisely} a duration $\tau$ in the
right half-space. Again, we cannot avoid employing a highly accurate meter. The result of our attempt is known from Sect. VII: we can neglect all but the singular in $\tau$ terms, thus obtaining
 \begin{eqnarray}\label{d8}
P(x,t|\tau)=|\psi_{\infty}(x,t)|^2 \delta(\tau),
\end{eqnarray}
which corresponds to the wave packet reflected as if by an infinite wall, at the meter with only zero readings.
In the light of what was said above, this is hardly surprising. Classically, one can arrange 
a meter which would not perturb the measured system. One can also set the meter in such a way that it would act on the system, and then correctly measure the variable for the system perturbed by the very measurement \cite{DS2}. Quantally, the first option is not available since it is necessary to destroy interference. Our ideal measurement of is, in this sense, correct: in order to measure $\tau$ to a great accuracy, the meter would need to exert a force which wouldn't let the particle enter the region, and then confirm the zero result.
In a similar way, we can analyse the 'softer' measurements involving different shapes and widths of the function $G$. For example a detailed analysis of Aharonov's 'weak measurements' \cite{W1}-\cite{W3} can be found in \cite{DS7}.

Equally important are the restrictions which the above principle puts on what one can use the fine-grained amplitudes for. For example, simply summing $\tilde{\Phi} (x,t|\tau)$ over a certain range of $\tau$, squaring the modulus of the result, and declaring it the probability to have $\tau$ within a range is dangerous. The authors of \cite{HY} have tried separating the amplitudes into just two classes, those for which $\tau$ is zero, and those for which it is not, 
so that the two amplitudes are given by
$A(x,t|\tau=0)=-\psi_{0}(x,t)$ and  $A(x,t|\tau\ne 0)=\int d\tau [ \tilde{\Phi} (x,t|\tau)+ \tilde{\Phi} (-x,t|\tau)]=\psi_{0}(x,t)+\psi_{0}(-x,t)$. Now the probability to have entered the right half-space appears to have the value 
 \begin{eqnarray}\label{d9}
P(t|\tau\ne 0)\equiv \int dx|A(x,t|\tau)|^2 =\\ \nonumber
\int dx|\psi_{\infty}(x,t|\tau)|^2 + \int dx|\psi_0(x,t|\tau)|^2= 2, 
\end{eqnarray}
which, as the authors of \cite{HY} pointed out, must be wrong.
To find a reason for this discrepancy we revisit the measurement as defined by Eqs.(\ref{b2}) and (\ref{c3}).
Apparently, no initial state of the meter $G(f)$ effects the coarse-graining of the 
fine-grained amplitudes into these two classes. We, therefore, no longer have Eq.(\ref{bb5}), itself a consequence of unitarity of the system-meter evolution. Moreover, non-conservation of the number of particles in Eq.(\ref{d9}) suggests that the probabilities $P(t|\tau\ne 0)$ and $P(t|\tau= 0)$
are not measurable by any other scheme, short of injecting more particles into the system.
\section{The probability not to be absorbed in the right half-space}
Another case discussed in the Ref. \cite{HY} is absorption of a particle by an optical potential 
confined to the right half space, 
 \begin{eqnarray}\label{e1}
U(x)=iU\theta(x). 
\end{eqnarray}
As before, we consider a wave packet incident on the absorbing potential from the left,
and a time $t$ so large that a free pulse would be fully contained to the right of the origin $x=0$.
Following \cite{HY} we wish to approximate the probability not to enter the right half space with the probability not to be absorbed by $U(x)$.
The problem is easily solved by the technique of the previous Section. The amplitude not to be absorbed after travelling along a Feynman path $x(t')$ is obviously $\exp\{iS_0[x(t')] -Ut_{\Omega}[x(t')]\}$, with $S_0$ denoting the free-particle action. The amplitude to arrive in $x$ by travelling along all paths satisfying $t_{\Omega}[x(t')]=\tau$ is, therefore, $\Psi(x,t|\tau)\exp(-U\tau)$, and the amplitude to arrive there at all is 
  \begin{eqnarray}\label{e2}
\psi_U(x,t)=\int_0^t \exp(-U\tau')\Phi(x,t|\tau')d\tau',
\end{eqnarray}
where $\Phi(x,t|\tau')$ is given by Eq.(\ref{d6}).
This is very similar to the amplitude for obtaining a zero reading if one measures the duration spent in the right half space by a meter whose initial state $G(\tau)$ is $\theta(\tau)\exp(-U\tau)$.
We, therefore, already know what will happen if one increases absorption $U$ in order to eliminate the particles which have entered the right half-space, and then managed to survive until $t$.
Since $\int d\tau' \theta(\tau')\exp(-U\tau')=1/U$, which vanishes as $U \to \infty$, the contributions from the smooth part of $\Phi(x,t|\tau')$ will vanish, leaving us again with the particle fully reflected from the origin,
  \begin{eqnarray}\label{e2a}
\psi_U(x,t)\approx\psi_\infty(x,t), \q as\q U\to\infty.
\end{eqnarray}
This result is equivalent to two complimentary statements. (a) Restricting an evolution to the Feynman paths which do not enter a spacial region leads to perfect reflection from the region's boundary. (b) Such a restriction can be achieved by introducing a large absorbing potential, the physical cause of the reflection. 
\section{The time of crossing into the right half-space for the first time}
The third case considered in \cite{HY} involves controlling the first time a system enters a given region of space $\Omega$ or, more generally, a certain sub-space of its Hilbert space. For a particle of a mass $M$ in one dimension, we construct a functional whose value gives the time a Feynman path enters $\Omega$ for the first time. 
 \begin{eqnarray}\label{f1}
\Theta_{\Omega}[x(\.)]=lim_{\gamma \to \infty }\int_0^tdt'\exp\{-\gamma t_{\Omega}(t',[x(\.)])\}
\end{eqnarray}
where $t_{\Omega}([x(t'),t)]$ is the traversal time functional defined earlier in Eq.(\ref{d1}).
The functional adds up $dt$'s for as long as $\exp\{-\gamma t_{\Omega}[x(t)]$ is not zero,  i.e., 
for as long as the path had made no incursion into $\Omega$, and marks the moment the border of $\Omega$ is crossed for the first time. If the path originates from inside $\Omega$ at $t=0$, the value of $\Theta_{\Omega}[x(t)]$ is set to zero. If a path has not yet visited $\Omega$ by the time $t$, the value is set to $t$. In this way, every Feynman path is labeled by its first crossing time, and we can re-arrange the paths into the classes, as was done in Sect..
For the amplitude to first cross into $\Omega=[0,\infty)$ at a time $\tau$,  $0\le 
\tau \le t$ we have
 \begin{eqnarray}\label{f2}
\Phi(x,t|\tau)=
\int dx' \la x'|\psi_I\ra \q\q\q\q\q\\ \nonumber
\int_{x(0)=x'}^{x(t)=x} Dx \exp\{iS_0[x(\.)\}\delta(\Theta_{\Omega}[x(\.)]-\tau)
\end{eqnarray}
where the path integration is over the paths starting at $t=0$ in $x'$ and ending in $x$ at the time $t$. The first crossing time expansion has been derived by many authors \cite{FP1}-\cite{FP5}, and in the Appendix we offer yet another one based on the direct evaluation of the restricted path integral in
(\ref{f2}). For a wave packet (\ref{d4a}) approaching the origin from the left, from Eq.(\ref{Ap5}) we have
 \begin{eqnarray}\label{f3}
\Phi(x,t|\tau)= \psi_{\infty}(x,t)\delta(\tau-t) \q\q\q\q\q\\ \nonumber
+(i/2M)\theta(\tau)\theta(t-\tau)K(x,0,t-\tau)\partial_x \psi_{\infty}(0,\tau),
\end{eqnarray}
where, as before, $\psi_{\infty}(x,t)$ is the wave packet scattered by an infinite wall at $x=0$.
We see that initially [$0_-$ stands for $lim_{\e\to 0}(0-\e)$]
 \begin{eqnarray}\label{f4}
\Phi(x,0_-|\tau)=\psi_I(x)\delta(\tau),
\end{eqnarray}
and, using Eq.(\ref{Ap4}), find the equation of motion, 
 \begin{eqnarray}\label{f5}
i\partial_t \Phi(x,t|\tau)= 
\h \Phi(x,t|\tau)\q\q\q\q\q\q\q\q\q\q\\ \nonumber
-i\partial_{\tau}[\delta(\tau-t)\psi_{\infty}(x,t)\theta(t)].\q\q\q\q\q\q\q
\end{eqnarray}
Integrating Eq.(\ref{f5}) over $\tau$ shows that
 \begin{eqnarray}\label{f6}
\int d\tau \Phi(x,t|\tau) = \psi(x,t),
\end{eqnarray}
which, together with Eq.(\ref{f3}) gives the standard first crossing time expansion, used for example in \cite{FP4}. This completes the first task outlined in Sect. VI, that is, defining the quantity of interest.

We do, however, fail in the second task, in specifying a meter for the first crossing time. Indeed, the Eq.(\ref{f5}) does not look like a SE describing interaction between  two quantum systems. It is non-homogenous, with the source term fully determined by the evolution in the left half-space with an infinite wall at the origin. We cannot even guarantee that it conserves the probability. Indeed constricting with the help of Eq.(\ref{b3}) a square integrable solution to represent the particle and a potential meter,
 \begin{eqnarray}\label{f7}
 \Psi(x,0_-|\tau)=\psi_I(x)G(\tau), \q \int | \Psi(x,0_-|\tau)|^2 d\tau dx =1,\q\q
\end{eqnarray}
and using Eq.(\ref{f5}) to evaluate the rate of change of $P(t)= \int | \Psi(x,t|\tau)|^2 d\tau dx$, 
we find
 \begin{eqnarray}\label{f8}
\frac{dP(t)}{dt}=\q\q\q\q\q\q\q\q\q\q\q\q\q\q\q\q\q\q\\ \nonumber\\ \nonumber
2 \mbox{Re}[\int_0^{\infty} d\tau \partial_{\tau}G(\tau-t)
\int dx \psi_{\infty}(x,t) \Psi^*(x,t|\tau)].
\end{eqnarray}
It is unlikely that the r.h.s. of Eq.(\ref{f8}) vanishes identically, and we abandon our attempts to find probabilities for the first crossing time (\ref{f1}), just as we promised in we would, in the second passage of Section ...
\section{Conclusions and discussion}
In summary, assigning probabilities to interfering alternatives implies destruction of coherence between the alternatives. 
Destruction of interference [e.g., conversion of a pure state (\ref{c2}) into a statistical mixture (\ref{c4})]
 is a physical process, and must be executed by a physical agent, which we call a meter.
This puts serious restrictions on the probabilities we may construct.
\newline
Firstly, such a meter must exist. Precise type of the interaction, required to destroy the interference, is prescribed by the property of Feynman paths we wish to control. It may happen that it does not correspond to a coupling  which is physically acceptable.
\newline
Secondly, even if the meter exists, it must be capable of effecting the desired separation of Feynman paths into classes. This requires finding an acceptable initial state for the meter.
This may not always be possible.
\newline
Thirdly, a measurement must be classified according to its accuracy, determined by the width and shape of the initial meter's state.
The range of possible measurements stretches for highly inaccurate 'weak' measurements to highly accurate 'ideal' ones. 
\newline
A simple analysis of Sect. VIII shows that an ideal measurement of the type discussed in \cite{HY} may lead to a kind of a Zeno effect. 
This is neither an 'unphysical' result, nor a 'pitfall' of  a theory,  but a general quantum mechanical rule.
To even contemplate the accurate  value of a variable $F$, one must assume that the interference has been destroyed to the required degree, i.e., consider also the effects of an external meter.
The meter would then perturb the system, and yield a sharply defined value of $F$, which correctly describes this perturbed motion.
\newline
This 'Zeno effect' cannot be avoided completely.  It can, however, be mitigated, e.g.,  by requesting less information about $F$, and reducing the accuracy of the measurement $\Delta f$.
The authors of Refs. \cite{HY} and \cite{KICK} chose to consider an ideal measurement of a quantity obtained
by replacing the integral (\ref{b1}) with a discrete sum (\ref{cb1}). With this, the system is allowed to evolve freely between $t_n$ and $t_{n-1}$, and is not reduced to the Zeno-like evolution if $\e$ is kept sufficiently large.  In general, one is lead to consider finite-time measurements of different quantities to different accuracies, varying $\Delta f$ and $\e$, to achieve a desired ratio 
between the information obtained and the perturbation incurred. 

We illustrate the above with a brief review of the examples considered in Ref.\cite{HY}
\newline 
(a) The probability for a free particle not to enter the right half-space equals the probability to spend there a zero net duration. This duration is represented by the traversal time functional, the relevant meter exists as a continuous version of the  Larmor clock \cite{}. An accurate clock prevents the particle from entering the region and, under the circumstances, the probability not to enter is unity.
\newline 
(b) The probability amplitude to enter the right half-space cannot be defined as the net probability to spend there any duration other than zero. Even though the meter (Larmor clock) exists, it cannot be set up to separate the Feynman paths into these two classes. One may be confident that this cannot be done by any other means either, since thus defined probabilities do not add up to unity.
\newline
(c) The probability not to be absorbed in the right half-space is similar to the probability not to enter it, and tends to unity as the magnitude of the absorbing potential increases.
This is a different way of saying that an infinite absorbing potential must reflect all incoming particles.
\newline
(d) We found no meaningful probabilities associated with the first crossing time amplitudes (\ref{f3}), as no physical meter of the type discussed here can be realised. To identify the time of  the  first crossing, one needs the record  the particle's past.
 For this reason, even classically, we cannot construct a single pointer which would stop once the particle first crosses into the region of interest. 

Finally, we note that in all above cases we have failed to arrive at the classical limit. This may cause some concern \cite{HY} 
Where a meter for our finite-time measurement exists, the remedy is simple \cite{DS2}. We do not improve accuracy indefinitely, but stop while $\Delta f$ exceeds the width of the stationary region of the fine-grained distribution in Eq. (\ref{b2}). This region occurs around the classical value of $F$, $f_{cl}$, and is very narrow  if the system is nearly classical. With the accuracy chosen high yet finite, only this region contributes to the integral (\ref{b3})
  Thus the pointer always points at $f_{cl}$, and we can replace the SE by the classical equations of motion.

Where no meter exists, the situation appears to be more difficult. Classically, one can always define the first crossing time, seemingly with no reference to meters or measurements. 
This is not quite so, as one always has at his disposal a classical trajectory $\bar{x}(t)$ from which all other quantities, including the first crossing time, can be derived. From quantum mechanical point of view, $\bar{x}(t)$ comprises the results of measuring the particle's positions at all times. This suggests that the first crossing time should be determined in a continuous quantum measurement, yielding a sequence of particle's positions $x(t)$, and then evaluating the functional (\ref{f1}) on this 'real', rather than virtual, trajectory. A more detailed analysis will be given in our future work \cite{DS8}.
\acknowledgements
We acknowledge support of the Basque Government (Grant No. IT-472-10), and the Ministry of Science and Innovation of Spain (Grant No. FIS2009-12773-C02-01). I am also grateful to J.N.L. Connor for bringing the issue to my attention.
\section{Appendix}
Consider the path integral in Eq.(\ref{f2}) keeping, for the moment $\gamma$ finite. Using the identity $\delta(z)=(2\pi)^{-1/2}\int \exp(i\lambda z) d\lambda$ w can rewrite it as
 \begin{eqnarray}\label{Ap1}
K(x,x',t|\tau)=(2\pi)^{-1/2}\int d\lambda 
\exp(i\lambda \tau)\times \q\q\q\q\q\\ \nonumber
\int_{x(0)=x'}^{x(t)=x} Dx \exp\{i\int_0^t dt' [ L_0 
-\lambda \exp(-\gamma\int_0^{t'} \theta_{\Omega}(x(t'')dt'' ]\}
\end{eqnarray}
where $L(x,\dot{x})$ is the particle's Lagrangian. Expanding the second exponential we find the amplitude for a path $x(t')$ to be 
 \begin{eqnarray}\label{Ap2}
A'[x(t')]=\exp(i\int_0^t  Ldt')\{1+\sum_{n=1}^{\infty}(-i\lambda)^n\times\q\q\q\\ \nonumber
\int_0^{t}dt_{1}\int_0^{t_{1}}dt_{2}..\int_0^{t_{n-1}}dt_n
\exp[-\gamma \sum_{m=1}^n \int_0^{t_m} \theta_{\Omega}(x(t'))dt'\}.
\end{eqnarray}
The first term in the curly brackets corresponds to free motion. The integrand of the $n$-th term corresponds to the particle moving in a time depended optical potential.
For $0\le t' <t_{n-1}$ we have $n\gamma\theta_{\omega(x)}$, for $t_{n-1}\le t' <t_{n-2}$ the potential is reduced to $(n-1)\gamma\theta_{\omega(x)}$, and so on. For $t\le t'<t_1$ the absorbing potential is turned off. As we send $\gamma \to \infty$,
the distinction between, say, the terms containing $n\gamma$ and $(n-1)\gamma$ disappears, and 
the particle moves in an infinite absorbing potential until $t_n$, after which it is switched off.
Integrals $\int_0^{t_{1}}dt_{2}...\int_0^{t_{n-1}}dt_n$ can now be evaluated to yield 
$t_n^{n-1}/(n-1)!$,
and summing over $n$ we have
 \begin{eqnarray}\label{Ap3}
A'[x(t')]=\exp(i\int_0^t  Ldt')- \q\q\q\q\q\\ \nonumber
i\lambda\int_0^t dt_1\exp(-i\lambda t_1)\exp(i\int_{t_1}^t Ldt'+\int_0^{t_1}L_{\infty}dt'),
\end{eqnarray}
where $L_{\infty}=\lim_{\gamma \to \infty }L -\gamma \theta_{\Omega}(x)$ is the Lagrangian
with an infinite absorbing potential introduced in the right half-space. Inserting Eq.(\ref{Ap3}) into (\ref{Ap1}) yields the net amplitude on all paths connecting $x'$ and $x$, and first crossing the origin $x=0$ at the time $\tau$
 \begin{eqnarray}\label{Ap4}
K(x,x',t|\tau)= K(x,x',t)\delta(\tau)-\q\q\q\q\q\\ \nonumber
-\partial_{\tau}\int dx''K(x,x'',t-\tau)K_{\infty}(x'',x',\tau) dx''
\end{eqnarray}
where $K_{\infty}$ and $K$ are the propagators with and without the infinite absorbing potential 
in the right half space, i.e., $K(x,x',t-\tau)\equiv \theta(t-\tau) \la x |\exp(-i\h t)|x'\ra$ and
$K_{\infty}(x,x',\tau)\equiv \theta(\tau) \la x |\exp(-i\h \tau)|x'\ra$, where $\h_{\infty}\equiv 
lim_{\gamma \to \infty}\h -\gamma \theta(x)$
 . Note that
the infinite absorbing potential is equivalent to an infinite potential wall introduced at $x=0$, so that
$$K_{\infty}(x,x',t) \equiv 0\q \mbox{for} \q x,x'\ge 0.$$
Differentiating the product in (\ref{Ap4}) and noting that $\h \theta(x)-\theta(x)\h_{\infty}=
-(1/2M)[\partial_x^2, \theta(x)]= (1/2M)\delta(x) \partial_x$, we rewrite
Eq.(\ref{Ap4}) as
 \begin{eqnarray}\label{Ap5} \nonumber
K(x,x',t|\tau)= K(x,x',t)\delta(\tau)\theta(x)+ K_{\infty}(x,x',t)\delta(\tau-t)+\\
(i/2M)K(x,0,t-\tau)\partial_{x''}K_{\infty}(x''=0,x',\tau),\q\q\q\q\q\
\end{eqnarray}
which is the standard form of the first crossing time amplitude for a particle travelling between $x'$ and $x$ \cite{FP1}-\cite{FP5}.

 
\end{document}